# Windstorm Economic Impacts on the Spanish Resilience: A Machine Learning Real-Data Approach

Matheus Puime Pedra
*Industrial Management Department*
*TECNUN, University of Navarra, Donostia, Spain*
mpuime@unav.es

Josune Hernantes
*Industrial Management Department*
*TECNUN, University of Navarra, Donostia, Spain*
jhernantes@unav.es

Leire Casals
*TECNUN*
*University of Navarra, Donostia, Spain*

Leire Labaka
*Industrial Management Department*
*TECNUN, University of Navarra, Donostia, Spain*
llabaka@unav.es

*Abstract*—Climate change-associated disasters have become a significant concern, principally when affecting urban areas. Assessing these regions' resilience to strengthen their disaster management is crucial, especially in the areas vulnerable to windstorms, one of Spain's most critical disasters. Smart cities and machine learning offer promising solutions to manage disasters, but accurately estimating economic losses from windstorms can be difficult due to the unique characteristics of each region and limited data. This study proposes utilizing ML classification models to enhance disaster resilience by analyzing publicly available data on windstorms in the Spanish areas. This approach can help decision-makers make informed decisions regarding preparedness and mitigation actions, ultimately creating a more resilient urban environment that can better withstand windstorms in the future.

## I. Introduction

In an era characterized by escalating urbanization and the frequency of natural disasters, it is imperative to fortify urban areas against these adversities [1]. Windstorms pose a significant threat to urban resilience in Spanish regions, being the second natural disaster that causes the most economic losses reported by insurance companies [2]. In this case, resilience refers to the ability of a region to prepare and plan for, absorb, recover from, and adapt to current and future stresses [3]. Smart cities with interconnected technologies and data-driven systems have immense disaster management and response potential. In addition, machine learning (ML) models can be used to discover intricate patterns and relationships and assist practitioners in their decision-making [1].

Different perspectives can be adopted when analyzing the effects of a windstorm disaster in a city or region, and one of them is the associated economic losses to a stress event. Economic losses can give an overview of the physical losses when a disaster occurs. However, estimating such losses is challenging due to the distinct characteristics of the disaster region and resilience [4], [5]. In addition, obtaining the data associated with natural disasters is challenging due to the lack of findability, accessibility, and interoperability [5], [6].

Several studies have attempted to analyze the resilience level of cities and regions by estimating economic losses. However, accurate data, proprietary information, and indicators related to resilience are often lacking, making it challenging to capture the essence of the analyzed region [4], [5], [6], [7]. Therefore, it is imperative to consider all these factors when adopting ML models to understand their economic effects when a disaster occurs in a region or city.

This study explores the crucial role of ML estimation models in bolstering disaster resilience management by utilizing publicly accessible data related to windstorm events and geographic and socioeconomic characteristics of Spanish regions. The suggested approach could offer decision-makers valuable resource information for making informed and proactive decisions on disaster preparedness and mitigation strategies. Ultimately, this could pave the way for Spanish cities to strengthen their resilience against the challenges posed by windstorms in the future.

## II. State of the Art

Smart City technologies contribute to disaster resilience, enhancing the region's disaster management, helping to be better prepared for disasters, predicting their effects, taking proactive measures, and developing post-disaster resilience operations [1]. Adopting the advantages of technology and innovation, such as ML models associated with reliable data, is paramount for promoting resilience in Smart Cities.

ML adoption in disaster management and the Smart City area has been essential to build resilience and assist practitioners [1]. With the ability to analyze and find patterns and tendencies in large volumes of distinct associated disaster data, ML can generate reliable and trustworthy information for preparedness and mitigation plans [8]. However, it is necessary to note that the current approaches report a lack of consistent, accurate, and real data for performing the analysis [4], [8]. This is associated with the fact that to reflect the current disaster scenario and the analyzed region characteristics, plenty of different data needs to be collected. In such cases, the necessary data are restricted or face some drawbacks [5].

Furthermore, some methods do not incorporate the features of disaster resilience when implementing machine learning models. This omission can be attributed to the data utilized in these models, which typically lacks any association with the concept of disaster resilience. By adopting this concept to define and collect the data used in the analysis, the dependability of the results can be improved, as the chosen data aligns with the resilience concept. Conversely, failing to incorporate a resilience-based approach in selecting the appropriate data for analysis can add to the complexity of practitioners' comprehension of the outputs. Embracing resilience can provide insights into the unique characteristics of disasters and their impact on a region or city [1], [9].

Distinct approaches tend to focus on this analysis principally by adopting the associated economic losses to detect the consequences of the analyzed disasters in a region. The work of [10] uses ML to predict flood probability in the Texas Gulf Coast region. An analysis of insurance claims and geospatial data generated a continuous flood hazard map, and the results showed that the random forest model predicted flooding with high accuracy. Another example is the work of [5], which proposes an economic loss forecasting system for storm surge disasters adopting ML and selecting key attributes. Following this approach, the work of [4] presents a ML model to predict earthquake casualty rate and economic losses implemented on the seismic loss dataset of mainland China to demonstrate its practicability and high predictive abilities.

The mentioned works reported the difficulty in collecting and using disaster event data, associating such data to the resilience concept and providing reliable method to estimate the associated economic disaster losses. Based on that, there is still some room for improvement in providing a framework to assist practitioners in disaster management tasks in cities and regions. This work presents a random forest classifier (RFC) associated with reliable and accurate data based on the definition of the resilience concept to estimate economic loss levels when a windstorm occurs in Spanish provinces.

## III. METHODOLOGY

The primary purpose of this research is to identify the nonlinear relationship between a set of features and the possible economic losses incurred during wind storms in Spanish regions. This section encompasses four key phases, namely: (1) data collection, (2) data pre-processing, (3) adoption of the ML algorithm to generate the classification model, and (4) performance evaluation. The research framework is shown in Fig. 1. A summary of each step is provided in the subsequent subsections.

### A. Data Collection

This paper analyzes 204 disaster events caused by wind storms that generated massive economic losses to regions and were reported by the Spanish Insurance Compensation Consortium (CCS) over a ten-year period (2013 to 2022). The collected data was classified into three categories to identify the occurrence of wind storm disaster events, the meteorological characteristics, and the associated resilience features. The content of each category, the associated features with the respective unit and temporal resolution are shown in Table 1.

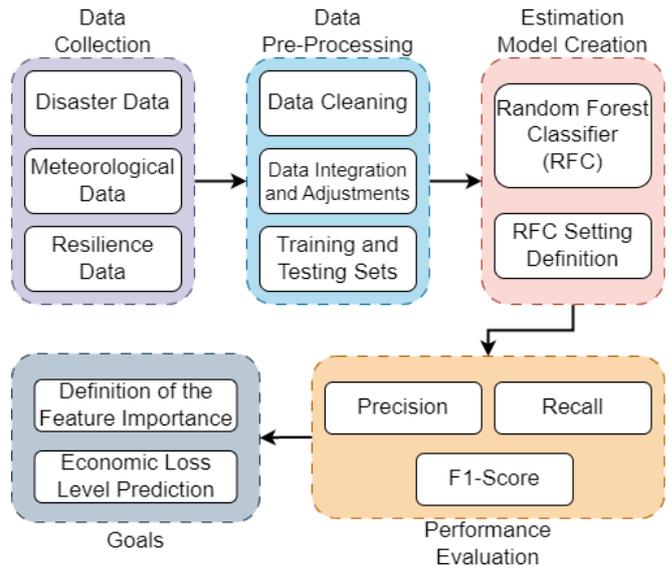

Fig. 1. The research framework of the prediction of the economic loss level

Initially, we collected data related to disasters and their associated events. These details were sourced from the CCS and encompassed all insurance claims information reported over a decade. The information included the province name where the event took place, the associated economic losses incurred in euros, and the date when the event started.

The meteorological category includes the markers linked to windstorm events. Our source for data extraction was the Spanish State Meteorological Agency's (AEMET) open data portal. The pertinent data includes wind speed in Kilometers per hour, precipitation in millimeters, and temperature in degree Celsius during the disaster episodes. Furthermore, we also incorporated the duration of the event in days, which was sourced from the CCS dataset.

To gauge a region's resilience, we have incorporated the Baseline Resilience Indicators for Resilience (BRIC) framework, which encompasses six key dimensions related to disaster resilience. These resilience subdomains cover various topics, including social, economic, community capital, institutional, housing and infrastructural, and environmental. According to [11], the BRIC framework is a valuable baseline for assessing disaster resilience in urban and rural areas.

We utilized BRIC resilience concepts to reflect the unique characteristics of the Spanish region impacted by wind storms. We gathered from the Spanish Statistical Office (INE) 11 features that align with the social, economic, infrastructure, and environmental resilience concepts. These features were readily available in open repositories and are at the province level. For social resilience, we incorporated median age, total population, population over 65, population under 15, the population between 16 and 64, and the percentages of Spanish and foreign populations. Economic resilience was measured by the rate of employment and unemployment, GDP per capita, average salary, household spending, and income. Infrastructure resilience was evaluated based on the number of affected systems and the length of the road network. Lastly, environmental resilience was assessed through data related to the total surface area of the province, agricultural surface area, and coastline length. A correlation matrix was created to analyze the association between the collected distinct features, as shown in Fig. 2. The results indicate that the social resilience data are correlated, confirming the classification presented with the BRIC framework. This correlation is also evident in the economic resilience variables. Regarding meteorological values, a correlation is present between temperature and wind variables, but there is no correlation between precipitation, season, and duration. Before performing the cleaning steps and using these features, we collected information on 756 windstorm disaster events.

TABLE I. DATA SOURCES AND CONTENTS

| Category[a] | | Feature | Unit | Temporal Resolution |
|---|---|---|---|---|
| Disaster | | Province name | String | – |
| | | Total economic losses | Euros | Event |
| | | Start date | Date | Event |
| Meteorological | | Duration | Days | Event |
| | | Avg. wind speed | Km/h | Event |
| | | Max wind speed | Km/h | Event |
| | | Precipitation | MM | Event |
| | | Avg. temperature | °C | Event |
| | | Max temperature | °C | Event |
| | | Min temperature | °C | Event |
| | | Season | Type | Event |
| Res. | Social | Median age | Age | 1 Year |
| | | Total population | Persons | 1 Year |
| | | Population over 65 | % | 1 Year |
| | | Population until 15 | % | 1 Year |
| | | Population between 16 and 64 | % | 1 Year |
| | | Population Spanish | % | 1 Year |
| | | Population foreign | % | 1 Year |
| | | Employment rate | % | 1 Year |
| | | Unemployment rate | % | 1 Year |
| | Eco. | PIB per capita | Euros | 1 Year |
| | | Spending household | Euros | 1 Year |
| | | Income household | Euros | 1 Year |
| | | Avg. Salary | Euros | 1 Year |
| | Infra. | Affected systems | Quantity | Event |
| | | Road | KM | 1 Year |
| | Env. | Total Surface | KM | 1 Year |
| | | Agricultural surface | Ha | 1 Year |
| | | Coast | KM | 1 Year |

[a.] The acronyms used in this Category column are defined as follows: Res. stands for Resilience, Eco. stands for Economic, Infra. stands for Infrastructure, and Env. stands for Environmental.

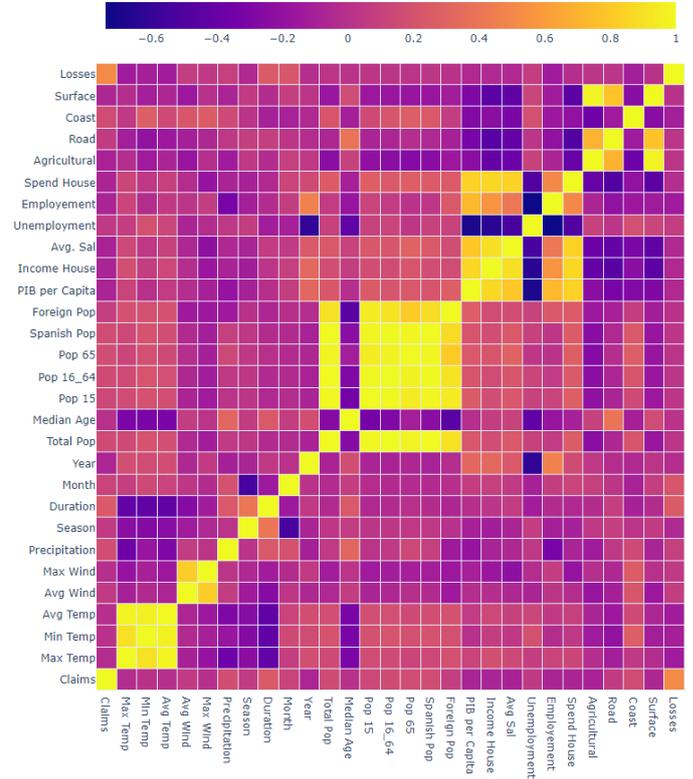

Fig. 2. Correlation matrix between the collected features

## B. Data Pre-Processing

After collecting all the features, a series of tasks were performed to avoid inconsistencies and inaccuracies that could potentially affect the results of the ML model. Given the vast temporal span of the collected data, preprocessing was necessary to adjust the data model for ML use. Three steps were followed to achieve this: cleaning the data, integrating and adjusting the final features, and generating the training and testing dataset. Due to the diversity of formats of the collected features, the Open Refine [12] open-source tool was employed to perform the cleaning steps. This step eliminated data inconsistencies such as null values and incorrect data formats.

Based on the cleaned data, a series of routines were conducted to integrate and adjust all the data into a single dataset. A script was developed to merge the meteorological features of the period of the event and the resilience features based on the year when the disaster occurred. It is essential to clarify that the resilience data are measured by year and province level, making it impossible to disaggregate at the level of days. To analyze the disaster events that caused massive economic losses for Spanish provinces, we performed an analysis of the windstorm economic losses' quartiles, focusing on selecting only the events that generated massive economic losses, which would be presented in the fourth quartile. As presented in Table 2, the fourth quartile was considered due to its distinction compared to the other quartiles. Based on that, 204 events were selected with a mean of 851,468.14 Euros ranging from 45,076 Euros to 19,298,377 Euros.

TABLE II. THE STANDARD FOR CLASSIFICATION OF WIND STORM LOSSES

| Quartile | Min and Max of Economic Losses in Euros | Economic Losses Mean |
|---|---|---|
| Q1 | 42 – 987 | 492.22 |
| Q2 | 1,014 – 5,860 | 2,700.58 |
| Q3 | 6,008 – 44,864 | 17,626.69 |
| Q4 | 45,076 – 19,298,377 | 851,468.14 |

Furthermore, the economic loss feature was transformed to classify its values better. This transformation was carried out by dividing the economic losses into three categories based on a thorough distribution analysis of the values. This approach allowed for a more precise and nuanced data classification, enabling better insights and decision-making. By excluding the events with low economic losses (events classified in Q1, Q2, and Q3 presented in Table 2) and categorizing the losses into three levels representing moderate (Level 1), severe (Level 2), and catastrophic (Level 3) economic losses, the results obtained can provide decision-makers with clear and concise information. Due to the small quantity of samples adopted in this approach, we avoided considering additional levels that may not represent a significant difference between them. Instead, the three-level category (shown in Table 3) offers more visual and direct results. Based on that, distinct disaster events were grouped based on the level of losses. Table 3 presents the economic loss level based on the distribution analysis, an overview of the range between the events, the mean, and the quantity for each level.

The last step of this phase was to divide the dataset into training and testing sets to be used for the machine learning task. Since only the disaster events that generated considerable economic losses were considered, the dataset is a small sample. Based on that, 75% of the samples were retained as the training set and 25% as the test set.

TABLE III. THE STANDARD FOR CLASSIFICATION OF WIND STORM LOSSES

| Economic Loss Level | Economic Losses in Euros | Economic Losses Mean | Quantity of Events |
|---|---|---|---|
| Level 1 | 45,076 – 110,451 | 71,233.98 | 68 |
| Level 2 | 112,089 – 575,431 | 256,001.41 | 68 |
| Level 3 | 630,242 – 19,298,377 | 2,227,168.98 | 68 |

*C. Estimation Model Creation*

For this work, we utilized the Random Forest Classifier (RFC), a popular non-parametric, supervised ML technique, to sort samples based on their highest predicted probability for each defined class. RFCs are formed by aggregating multiple random decision trees, which classify the input using the precision model of all the decision trees. In disaster management analysis, RFCs are often preferred over other decision trees due to their reliability, flexibility, and accuracy [10], [13], [14]. In our study, we used the RFC from the Scikit-learn Python library. In addition, we adopted a set of six parameters associated to the number of forest trees (n_estimators), the minimum number of samples needed to split an internal node (min_sample_split), the minimum number of samples required for a leaf node (min_sample_leaf), the number of features to consider when selecting the best split (max_features), the maximum depth of the tree (max_depth), and the function used to evaluate the quality of a split (criterion) [15].

We developed a script to identify the ideal combination of these settings, which relied on a pre-set of possible values for each parameter and automatically evaluated model performance. As part of our model performance, we adopted commonly used classification metrics such as Precision, Recall, and F1-score. Precision measures the accuracy ratio of positive predictions. Recall, on the other hand, measures the completeness of the positive predictions. Finally, the F1 score is used to assess the model's overall performance when there is a trade-off between Precision and Recall. A higher score of these three metrics indicates better model performance [15].

IV. RESULTS

The first run of the random forest classifier was adopted within the created dataset based on the Scikit-learn library's default definition. The model's precision was 0.764 based on the n_estimators equal to 100, min_sample_split of 2, min_sample_leaf value of 1, "sqrt" as the max_features, max_depth equal to none, and "gini" as the criterion. In addition, the recall results and the F1-score were equal to 0.76. These results represent an average performance of the model, and the setting should be improved to extract the best from the developed dataset.

Applying the developed script to find the best RFC settings for our dataset was perceived as an increase in the model's general performance, with accuracy reaching 0.823, recall, and F1-score achieving 0.82. The best RFC settings for the applied dataset were with the n_estimators equal to 1,135, min_sample_split of 5, min_sample_leaf value of 4, "auto" as the max_features, max_depth equal to 100, and "gini" as the criterion.

Analyzing the confusion matrix presented in Fig. 3 between the true and predicted labels, 42 samples were correctly predicted from the 51 samples from the testing dataset. However, it is perceived that some labels still have some inconsistencies, and in most cases, the labels are unpredictable to the class of a low level of economic losses compared to a high level of economic losses. The same result is perceived by generating the performance metrics based on the four-level economic losses represented in Table 4. It was perceived that the higher economic losses presented in level 3 obtained a higher score when compared to levels 1 and 2. This can be associated with the distinction between the amount of economic losses presented in levels 1 and 2.

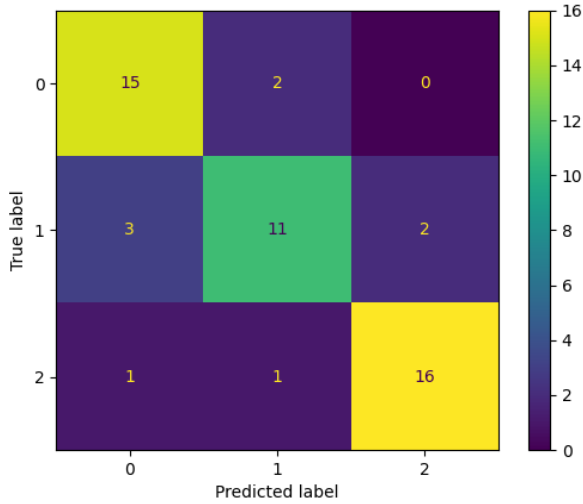

Fig. 3. Confusion matrix based on the generated outputs of the RFC

TABLE IV. ACCURACY OF THE BEST RFC MODEL

| Class | Precision | Recall | F1-score |
|---|---|---|---|
| Level 1 | 0.79 | 0.88 | 0.83 |
| Level 2 | 0.79 | 0.69 | 0.73 |
| Level 3 | 0.89 | 0.89 | 0.89 |

Fig. 4 presents the importance of each included feature. The most influential feature of the results was the number of affected systems during a wind storm event. This particular feature is closely associated with the infrastructure concept of resilience. It is a logical addition, considering the fact that the number of affected systems is directly proportional to the amount of reported economic losses. The second and the third most important features were the max wind velocity and the share of foreign population, respectively. The max wind is associated with the meteorological category and reflects the actual scenario because the higher is the wind intensity, the more harmful the disaster event will be. In addition, the max wind speed proves the model's reliability because it can be associated with the principal feature of measuring a windstorm disaster event. Another intriguing result of the model is the percentage of the foreign population, which could be linked to the notion that individuals who do not speak Spanish as their primary language may have greater difficulty comprehending government warning messages so that the losses can be higher.

By analysing the features based on their categories, we can see some interesting patterns that can assist stakeholders in understanding the results and adopting them in their disaster management analysis. It is notorious that meteorological features are the most influenced features in the dataset, which can be related to the already known definition of disaster management, where the higher the event magnitude, the higher the losses will be [16].

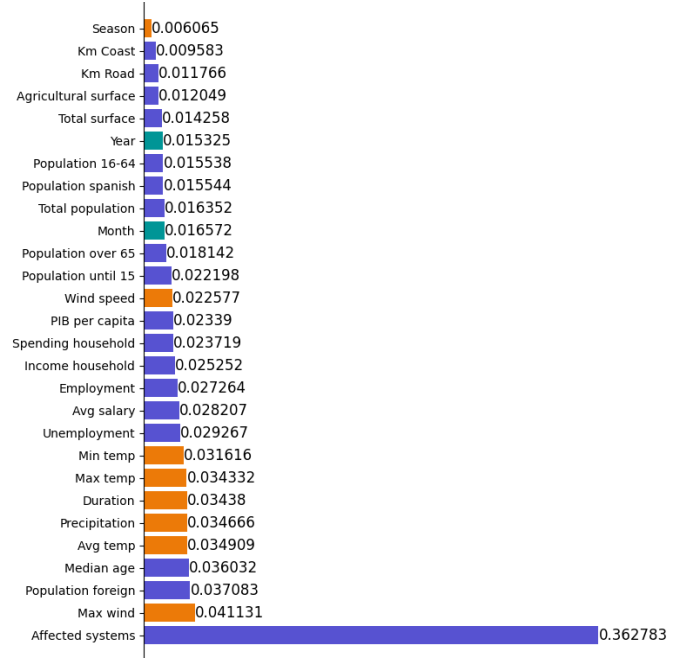

Fig. 4. Importance of each feature in the RFC

When considering the resilience concepts proposed by BRIC, the most influential factor in determining the resilience of infrastructures is the number of affected systems and the foreign population for social resilience. Regarding economic resilience, the unemployment rate is the most important, and for the resilience of the environment is the total surface feature. In general, the majority of resilience features hold similar importance in the model. The model's top five resilience features are the affected systems, the foreign population, the median age, the unemployment rate, and the average salary. These findings suggest, for example, that the unemployment rate of a Spanish province is more significant in the random forest classifier in estimating the level of economic loss than the employment rate. Similarly, the average salary impacts the model's outputs more than the PIB per capita. Finally, features related to environmental resilience were found to be the least important, with road length related to infrastructure resilience following closely behind.

V. DISCUSSION AND CONCLUSION

This work presented the application of an RFC to disaster management analysis based on accurate available data to assist stakeholders in understanding the current situation and being better prepared for the following stress events. The results demonstrated that applying RFC can effectively classify different types of disasters based on possible economic losses by using windstorm disaster events and meteorological and socioeconomic variables. By adopting a well-known classification model and appropriate settings, we generated a computationally efficient and scalable model that can be used to predict windstorm economic losses over relatively large regions. It is possible to incorporate the developed method with smart city early warning systems, other ML models, and sensor data to produce accurate predictions, allocate resources effectively, and create contingency plans for potential disasters [1].

As the feature's importance analysis indicates, almost all the adopted indicators considerably influence the model results. Indeed, the meteorological variables are the principal motor that indicates the obtained economic loss level in such cases, and the city stakeholders cannot change these variables. However, the resilience indicators category can be crucial for understanding and bolstering resilience in a region, particularly by focusing on resilience subdomains such as social, economic, infrastructure, and environmental aspects. By prioritizing these key elements, communities can be better equipped to withstand and bounce back from the effects of severe natural disasters. Our methodology can serve as a helpful starting point for urban practitioners looking to build disaster resilience management and bolster existing societal strengths, thereby enhancing preparedness, response, and recovery capabilities.

Another aspect of the developed approach is its adaptability to other regions based on the application of national analysis. As previously presented, the adopted data is about disasters that occurred in Spanish provinces so the national government can predict the following disaster economic losses and prioritize preparedness and mitigation actions in specific regions based on the region's characteristics. It is important to clarify that the following model adopts data from insurance companies, which may not encompass uninsured losses. Another issue is associated with analyzing the feature's importance, in such a case, it is important to highlight that a deeper analysis must be performed to understand the features' correlation to the obtained outputs. A further matter involves adopting a single classifier analysis, comparing our approach with other ML models could provide other analysis perspectives.

Finally, we eagerly anticipate further enhancements to our methodology, such as integrating more data on unique feature categories to enhance the BRIC resilience subdomains and performing predictions to estimate possible scenarios of economic losses. Moreover, we can expand our approach by incorporating an eXplainable Artificial Intelligence model to aid practitioners and city stakeholders in comprehending the resulting outputs. Additionally, we must consider climate change and socioeconomic patterns to estimate potential economic losses due to windstorms in the future.

## ACKNOWLEDGMENT

This work was supported by the Spanish Ministry of Science and Innovation under Grant PID2019-105414RA-C32; and under the 'Cátedra de Catástrofes', funded by Fundación AON.